\documentclass[sigconf]{acmart}

\setcopyright{none}
\renewcommand\footnotetextcopyrightpermission[1]{} 
\AtBeginDocument{%
  \providecommand\BibTeX{{%
    \normalfont B\kern-0.5em{\scshape i\kern-0.25em b}\kern-0.8em\TeX}}}

 \acmConference[IDEAIS'19]{Position paper}{October 1, 2019}{Position Paper}
\begin{document}

\title{IDEAIS: Smart Voice Assistants to Improve Interaction with SDIs}


\author{Miguel \'Angel Bernab\'e}
\affiliation{\institution{Universidad Nacional de Catamarca}}
\email{ma.bernabe@gmail.com}

\author{Jacinto Estima}
\affiliation{\institution{INESC-ID}}
\email{jacinto.estima@gmail.com}

\author{Mar\'ia Ester Gonz\'alez}
\affiliation{\institution{Universidad de Concepci\'on}}
\email{mariaesgonzalez@udec.cl}

\author{Carlos Granell}
\affiliation{\institution{Universitat Jaume I}}
\email{carlos.granell@uji.es}

\author{Carlos L\'opez-V\'azquez}
\affiliation{\institution{Universidad ORT Uruguay}}
\email{carloslopez@uni.ort.edu.uy}

\author{Miguel R. Luaces}
\affiliation{\institution{Universidade da Coru\~na, CITIC}}
\email{luaces@udc.es}

\author{Bruno Martins}
\affiliation{\institution{INESC-ID}}
\email{bruno.g.martins@ist.utl.pt}

\author{Daniela Moctezuma}
\affiliation{\institution{CentroGEO}}
\email{dmoctezuma@centrogeo.edu.mx}

\author{Diego Seco}
\affiliation{\institution{Universidad de Concepci\'on}}
\email{dseco@udec.cl}

\renewcommand{\shortauthors}{Bernab\'e, et al.}

\begin{abstract}
A critical goal, is that organizations and citizens can easily access the geographic information required for good governance. However, despite the costly efforts of governments to create and implement Spatial Data Infrastructures (SDIs), this goal is far from being achieved. This is partly due to the lack of usability of the geoportals through which the geographic information is accessed. In this position paper, we present IDEAIS, a research network composed of multiple Ibero-American partners to address this usability issue through the use of Intelligent Systems, in particular Smart Voice Assistants, to efficiently recover and access geographic information.
\end{abstract}

\maketitle

\section{Introduction}

The initial moments after a natural disaster, such as an earthquake or volcanic eruption, are key to an appropriate and rapid reaction of decision makers and first aid respondents, such as civil protection, military forces, governmental agencies, among others~\cite{cutter2003}. In these situations, having in place an intelligent system/assistant that could help them access all the relevant information related to the disaster in a few seconds could definitely be a game-changer to radically improve disaster and emergency management situations. With this motivation in mind, we present the IDEAIS network, a collaboration project funded by the CYTED program (\url{http://www.cyted.org}, Ibero-American Program of Science and Technology for Development) from 2019 to 2022. Although IDEAIS is an Ibero-American network, it is open for international collaboration.

Behind this motivation, there is a need to improve access to geographic information that is necessary for good governance \cite{vancauwenberghe2017}.
%which is an important aspect not only for the Ibero-American region but also at Global (GSDI~\cite{gsdi}), European (INSPIRE~\cite{inspire}), Pan-American (IPGH~\cite{ipgh}) and Latin American (GEOSUR~\cite{geosur}) scale. 
However, this goal is far from being achieved, despite the costly efforts of governments to develop Spatial Data Infrastructures (SDIs)~\cite{inspire-mid}. This is partly due to the lack of usability of geoportals and related geospatial services through which geographic information is accessed. Typically, decision makers near to the place of the event (i.e. mayors, local firefighters, local ambulances, etc.) are little or no experts at all in the use of SDIs (where the most complete geographic information is available), so it is essential to design an usable system to deliver the geographic information layers needed for each type of event. In the IDEAIS project, we propose the use of Intelligent Systems, paying close attention to recent technological developments related to the emergence of Intelligent Assistants to allow the development of Smart Voice Assistants for efficient recovery and access to geographic information.

Notwithstanding there is a significant improvement in voice assistants in various types of applications such as entertainment~\citep{Fernandes2019}, laboratory assistants~\citep{Austerjost2018}, and even in many daily life activities~\citep{Porcheron2018}; this kind of systems has been scarcely explored to access or manipulate geospatial resources accessible in SDIs. The above gives us a great opportunity to contribute in this area with the collaboration of this multidisciplinary research network.

As an illustrative example, consider a case in the context of civil protection: After the warning ``The Tungurahua has erupted'', a Smart Voice Assistant would allow decision makers the following immediate actions: i) discover that the Tungurahua is a volcano and that the expression ``has erupted'' indicates a state of emergency; ii) identify the geographic regions that may be affected; iii) find in the corresponding SDI nodes the necessary layers of geographic information and cartography in case of an eruption; iv) interact with underlying servers and geospatial services for geospatial data processing and integration, if required; and v) receive and redirect immediately responses to stakeholders (i.e. the one who sent the alert, emergencies, police, firefighters, etc.) near the volcano.

Although the technologies involved in this motivational example are to some extent in use in varied scenarios such as home automation and the development of Smart Cities \cite{degbelo2016}, there are some specific challenges that must be faced to incorporate such technologies for intelligent assistants in conjunction with the current state of SDIs and geospatial services and tools in our scenario. In Section~\ref{sect:architecture}, we first describe the conceptual architecture of a prototype of Intelligent Assistant as motivated earlier. Based on such reference architecture, we identify  in Section~\ref{sect:challenges} the main research challenges for our network. Finally, we specify the next steps to achieve our objective in Section~\ref{sect:outlook}.

\section{Architecture}\label{sect:architecture}

\begin{figure}[htbp!]
  \centering
  \includegraphics[width=0.78\linewidth]{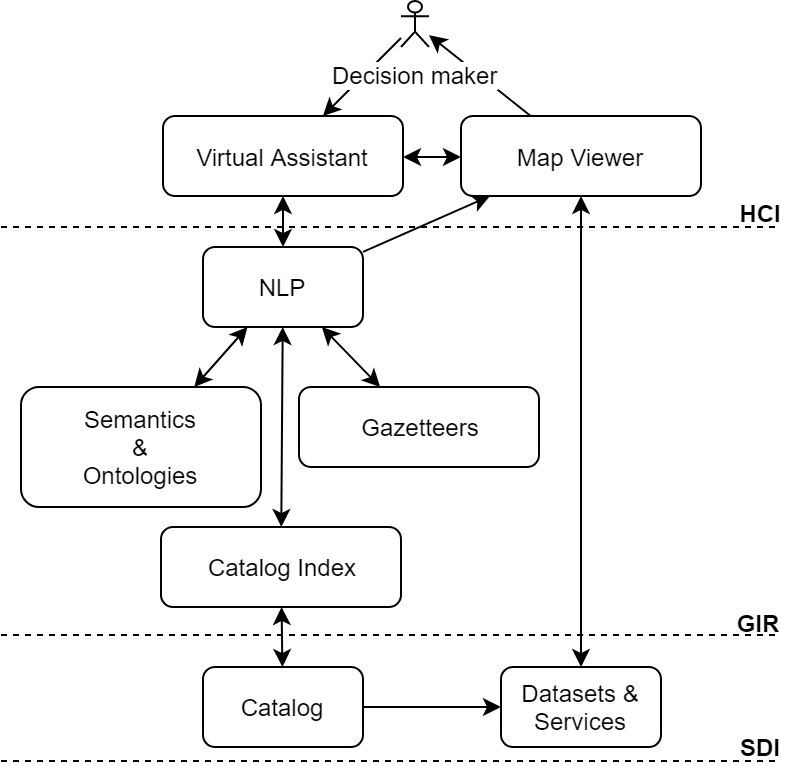}
  \caption{Proposed architecture}
  \label{fig:architecture}
\end{figure}

Figure~\ref{fig:architecture} provides a reference architecture for the prototype. The \emph{Virtual Assistant} component receives the user input and it uses the \emph{Natural Language Processing} (NLP) component to process the input and determine the geographic information queries that are needed. For this task, the NLP component uses the \emph{Semantics and Ontologies} component that models the disaster management domain, the \emph{Gazetteers} component that indexes geographic names, and the \emph{Catalog Index} component that indexes the catalog services of the SDI (represented by the \emph{Catalog} component in the architecture). Finally, the NLP component sends the geographic information queries to the \emph{Map Viewer} component that retrieves the data from the SDI datasets and services for data integration and fusion (represented by the \emph{Datasets \& Services} component). The components of the architecture can be divided into three layers: (i) a Human-Computer Interaction (HCI) package, (ii) a Geographical Information Retrieval (GIR) package, and (iii) a Spatial Data Infrastructure (SDI) package. These layers have been used in the research network to divide the project tasks into workpackages.

\section{Research Challenges}\label{sect:challenges}

In this section we identify and briefly describe the main research challenges to realize the Smart Voice Assistant and to achieve the objectives of our network. 

\begin{itemize}

    %\item \ldots
    %\item Research Challenge n
   \item \textit{Intelligent systems development}. The use of artificial intelligence techniques in geospatial data is a big challenge. In particular, the use of natural language processing methods to user communication or assistant, in natural language, to obtain information and interact with the SDI is a novel, current, and relevant research topic. These automatic question-answering (QA) assistants are becoming more popular in the research community and they have improved the voice interaction in very complex tasks~\citep{Callaghan2018}. 
    Nevertheless, its implementation with spatial data has been very little applied. Specifically, the two main challenges in the generation of intelligent virtual assistants to interact with SDI are i) the design of intelligent algorithms using natural language with voice or even text; and ii) creating algorithms that are able to reason and act with as less human interaction as possible is a hard task to solve.
    
    \item \textit{Communication between intelligent assistants and current standardization efforts regarding geographic information services}. Needless to say that Intelligent Assistants must still harness the vast array of available geospatial services and resources deployed in distributed SDI nodes to operate properly. Nevertheless, new usability-enhanced interaction modes (i.e. voice-based) require to rethink the traditional interfaces with the well-established Open Geospatial Consortium (OGC) standards-based services and resources such as those services for access geospatial data (e.g. WMS, WFS, WCS, and SensorThings) and for data processing and fusion (e.g. WPS), in order to find new ways (interfaces) to access and leveraging them without reinventing the wheel.
    
    \item \textit{Explore user experience challenges related with spatial data, information and knowledge required by decision makers in emergency management.} The characteristics of decision makers near the emergency, who may be little or no experts at all in the use of SDIs, require an evolution from the current approach of executing operations in SDIs to an approach driven by QA assistants. In addition, the study of legal guarantees and responsibilities in the use of geographic information is important in the emergency domain, in combination with the creation of a standardized knowledge base of terminology related to each type of event in natural disasters and emergencies. Finally, the International Standard Common Alerting Protocol (CAP) must be integrated, which also imposes some requirements regarding the user experience. 
    
    \item \textit{Design a scalable architecture and efficient components to retrieve all the necessary information within the strict time constraints of an state of emergency.} We will explore the use of a microservice-based architecture~\cite{sensors2019}, which requires a rethinking of the current service-oriented architectures of SDIs and associated communication protocols. As this is an ambitious goal, a new hybrid architecture that combines classic services and microservices, designed for some specific and critical tasks, will provide a gradual transition. For the implementation of the new microservices we will study the use of efficient indexes that combine spatial, temporal and textual dimensions. Finally, mere access to the required information is not enough, because data from different providers must be interoperable to accomplish the goal. Interoperability poses different requirements either in attributes, geometry or both. We will research in particular the geometric problem, which is manifested for example by parcels that lie in the sea, Points of Interest that are compatible with Google Earth imagery but not with official cartography of roads, and so on. Not much has been done on the topic, but we plan to set up a Geometrical Conflation Comparison Exercise to address the issue~\cite{usery2005}.

\end{itemize}

\section{Outlook}\label{sect:outlook}
By tackling the research challenges described above, IDEAIS network will contribute to advancing from traditional SDIs to next-generation Spatial Knowledge Infrastructures~\cite{ski2017}, in which the core element is knowledge instead of data or information. As we focus on rapid reaction after natural disasters, this is a crucial change to help decision makers who may not have the time or expertise to process and transform available information in SDIs to actionable knowledge.

%\section{Acknowledgements}
%Funded by CYTED: \emph{Programa Iberoamericano de Ciencia y Tencolog\'ia para el Desarrollo}, grant number 519RT0579.

\bibliographystyle{ACM-Reference-Format}
\bibliography{ideais}


\begin{thebibliography}{11}

%%% the Web address.

\ifx \showCODEN    \undefined \def \showCODEN     #1{\unskip}     \fi
\ifx \showDOI      \undefined \def \showDOI       #1{#1}\fi
\ifx \showISBNx    \undefined \def \showISBNx     #1{\unskip}     \fi
\ifx \showISBNxiii \undefined \def \showISBNxiii  #1{\unskip}     \fi
\ifx \showISSN     \undefined \def \showISSN      #1{\unskip}     \fi
\ifx \showLCCN     \undefined \def \showLCCN      #1{\unskip}     \fi
\ifx \shownote     \undefined \def \shownote      #1{#1}          \fi
\ifx \showarticletitle \undefined \def \showarticletitle #1{#1}   \fi
\ifx \showURL      \undefined \def \showURL       {\relax}        \fi
% invisible to TeX
\providecommand\bibfield[2]{#2}
\providecommand\bibinfo[2]{#2}
\providecommand\natexlab[1]{#1}
\providecommand\showeprint[2][]{arXiv:#2}

\bibitem[\protect\citeauthoryear{ANSORGE, Craglia, Fierens, HASENOHR, Jensen,
  LIHTENEGER, LUTZ, MILLOT, NUNES DE~LIMA, Roglia, Smits, and TOMAS}{ANSORGE
  et~al\mbox{.}}{2014}]%
        {inspire-mid}
\bibfield{author}{\bibinfo{person}{Christian ANSORGE}, \bibinfo{person}{Max
  Craglia}, \bibinfo{person}{Freddy Fierens}, \bibinfo{person}{Paul HASENOHR},
  \bibinfo{person}{Stefan Jensen}, \bibinfo{person}{Darja LIHTENEGER},
  \bibinfo{person}{Michael LUTZ}, \bibinfo{person}{Michel MILLOT},
  \bibinfo{person}{Maria NUNES DE~LIMA}, \bibinfo{person}{Elena Roglia},
  \bibinfo{person}{P Smits}, {and} \bibinfo{person}{Robert TOMAS}.}
  \bibinfo{year}{2014}\natexlab{}.
\newblock \bibinfo{booktitle}{\emph{Mid-term evaluation report on INSPIRE
  implementation}}.
\newblock
\showISBNx{978-92-9213-486-0}


\bibitem[\protect\citeauthoryear{Austerjost, Porr, Riedel, Geier, Becker,
  Scheper, Marquard, Lindner, and Beutel}{Austerjost et~al\mbox{.}}{2018}]%
        {Austerjost2018}
\bibfield{author}{\bibinfo{person}{Jonas Austerjost}, \bibinfo{person}{Marc
  Porr}, \bibinfo{person}{Noah Riedel}, \bibinfo{person}{Dominik Geier},
  \bibinfo{person}{Thomas Becker}, \bibinfo{person}{Thomas Scheper},
  \bibinfo{person}{Daniel Marquard}, \bibinfo{person}{Patrick Lindner}, {and}
  \bibinfo{person}{Sascha Beutel}.} \bibinfo{year}{2018}\natexlab{}.
\newblock \showarticletitle{Introducing a Virtual Assistant to the Lab: A Voice
  User Interface for the Intuitive Control of Laboratory Instruments}.
\newblock \bibinfo{journal}{\emph{SLAS TECHNOLOGY: Translating Life Sciences
  Innovation}} \bibinfo{volume}{23}, \bibinfo{number}{5}
  (\bibinfo{year}{2018}), \bibinfo{pages}{476--482}.
\newblock
\urldef\tempurl%
\url{https://doi.org/10.1177/2472630318788040}
\showDOI{\tempurl}
\newblock
\shownote{PMID: 30021077.}


\bibitem[\protect\citeauthoryear{Callaghan, Putinelu, Ball, Salillas, Vannier,
  Egu{\'i}luz, and McShane}{Callaghan et~al\mbox{.}}{2018}]%
        {Callaghan2018}
\bibfield{author}{\bibinfo{person}{Michael~James Callaghan},
  \bibinfo{person}{Victor~Bogdan Putinelu}, \bibinfo{person}{Jeremy Ball},
  \bibinfo{person}{Jorge~Caballero Salillas}, \bibinfo{person}{Thibault
  Vannier}, \bibinfo{person}{Augusto~Gomez Egu{\'i}luz}, {and}
  \bibinfo{person}{Niall McShane}.} \bibinfo{year}{2018}\natexlab{}.
\newblock \showarticletitle{Practical Use of Virtual Assistants and Voice User
  Interfaces in Engineering Laboratories}. In \bibinfo{booktitle}{\emph{Online
  Engineering {\&} Internet of Things}}. \bibinfo{publisher}{Springer},
  \bibinfo{pages}{660--671}.
\newblock
\showISBNx{978-3-319-64352-6}


\bibitem[\protect\citeauthoryear{Cutter}{Cutter}{2003}]%
        {cutter2003}
\bibfield{author}{\bibinfo{person}{Susan~L Cutter}.}
  \bibinfo{year}{2003}\natexlab{}.
\newblock \showarticletitle{GI science, disasters, and emergency management}.
\newblock \bibinfo{journal}{\emph{Transactions in GIS}} \bibinfo{volume}{7},
  \bibinfo{number}{4} (\bibinfo{year}{2003}), \bibinfo{pages}{439--446}.
\newblock


\bibitem[\protect\citeauthoryear{Degbelo, Granell, Trilles, Bhattacharya,
  Casteleyn, and Kray}{Degbelo et~al\mbox{.}}{2016}]%
        {degbelo2016}
\bibfield{author}{\bibinfo{person}{Auriol Degbelo}, \bibinfo{person}{Carlos
  Granell}, \bibinfo{person}{Sergio Trilles}, \bibinfo{person}{Devanjan
  Bhattacharya}, \bibinfo{person}{Sven Casteleyn}, {and}
  \bibinfo{person}{Christian Kray}.} \bibinfo{year}{2016}\natexlab{}.
\newblock \showarticletitle{Opening up Smart Cities: Citizen-Centric Challenges
  and Opportunities from GIScience}.
\newblock \bibinfo{journal}{\emph{ISPRS International Journal of
  Geo-Information}} \bibinfo{volume}{5}, \bibinfo{number}{2}
  (\bibinfo{year}{2016}), \bibinfo{pages}{16}.
\newblock
\showISSN{2220-9964}
\urldef\tempurl%
\url{https://doi.org/10.3390/ijgi5020016}
\showDOI{\tempurl}


\bibitem[\protect\citeauthoryear{Duckham, Arnold, Armstrong, McMeekin, and
  Mottolini}{Duckham et~al\mbox{.}}{2017}]%
        {ski2017}
\bibfield{author}{\bibinfo{person}{Matthew Duckham}, \bibinfo{person}{Lesley
  Arnold}, \bibinfo{person}{Kylie Armstrong}, \bibinfo{person}{David McMeekin},
  {and} \bibinfo{person}{Darren Mottolini}.} \bibinfo{year}{2017}\natexlab{}.
\newblock \bibinfo{booktitle}{\emph{{White Paper: Towards a Spatial Knowledge
  Infrastructure}}}.
\newblock \bibinfo{type}{{T}echnical {R}eport}.
  \bibinfo{institution}{{Australia and New Zealand Cooperative Research Centre
  for Spatial Information (CRCSI)}}.
\newblock


\bibitem[\protect\citeauthoryear{Fernandes, Abreu, Almeida, and
  Santos}{Fernandes et~al\mbox{.}}{2019}]%
        {Fernandes2019}
\bibfield{author}{\bibinfo{person}{S{\'i}lvia Fernandes},
  \bibinfo{person}{Jorge Abreu}, \bibinfo{person}{Pedro Almeida}, {and}
  \bibinfo{person}{Rita Santos}.} \bibinfo{year}{2019}\natexlab{}.
\newblock \showarticletitle{A Review of Voice User Interfaces for Interactive
  TV}. In \bibinfo{booktitle}{\emph{Applications and Usability of Interactive
  TV}}. \bibinfo{publisher}{Springer International Publishing},
  \bibinfo{address}{Cham}, \bibinfo{pages}{115--128}.
\newblock
\showISBNx{978-3-030-23862-9}


\bibitem[\protect\citeauthoryear{Li, Seco, and Rodr{\'{\i}}guez}{Li
  et~al\mbox{.}}{2019}]%
        {sensors2019}
\bibfield{author}{\bibinfo{person}{Zheng Li}, \bibinfo{person}{Diego Seco},
  {and} \bibinfo{person}{Alexis Eloy~S{\'{a}}nchez Rodr{\'{\i}}guez}.}
  \bibinfo{year}{2019}\natexlab{}.
\newblock \showarticletitle{Microservice-Oriented Platform for Internet of Big
  Data Analytics: {A} Proof of Concept}.
\newblock \bibinfo{journal}{\emph{Sensors}} \bibinfo{volume}{19},
  \bibinfo{number}{5} (\bibinfo{year}{2019}), \bibinfo{pages}{1134}.
\newblock


\bibitem[\protect\citeauthoryear{Porcheron, Fischer, Reeves, and
  Sharples}{Porcheron et~al\mbox{.}}{2018}]%
        {Porcheron2018}
\bibfield{author}{\bibinfo{person}{Martin Porcheron}, \bibinfo{person}{Joel~E.
  Fischer}, \bibinfo{person}{Stuart Reeves}, {and} \bibinfo{person}{Sarah
  Sharples}.} \bibinfo{year}{2018}\natexlab{}.
\newblock \showarticletitle{Voice Interfaces in Everyday Life}. In
  \bibinfo{booktitle}{\emph{Proceedings of the 2018 CHI Conference on Human
  Factors in Computing Systems}} \emph{(\bibinfo{series}{CHI '18})}.
  \bibinfo{publisher}{ACM}, Article \bibinfo{articleno}{640},
  \bibinfo{numpages}{12}~pages.
\newblock


\bibitem[\protect\citeauthoryear{Usery, Finn, and Starbuck}{Usery
  et~al\mbox{.}}{2005}]%
        {usery2005}
\bibfield{author}{\bibinfo{person}{E~L Usery}, \bibinfo{person}{M~P Finn},
  {and} \bibinfo{person}{M Starbuck}.} \bibinfo{year}{2005}\natexlab{}.
\newblock \showarticletitle{Integrating data layers to support The National Map
  of the United States}. In \bibinfo{booktitle}{\emph{Proceedings of the
  International Cartographic Conference}}.
\newblock


\bibitem[\protect\citeauthoryear{Vancauwenberghe and van
  Loenen}{Vancauwenberghe and van Loenen}{2017}]%
        {vancauwenberghe2017}
\bibfield{author}{\bibinfo{person}{Glenn Vancauwenberghe} {and}
  \bibinfo{person}{Bastiaan van Loenen}.} \bibinfo{year}{2017}\natexlab{}.
\newblock \showarticletitle{Governance of open spatial data infrastructures in
  Europe}.
\newblock In \bibinfo{booktitle}{\emph{The social dynamics of open data}}.
  \bibinfo{publisher}{African Minds}, Chapter~4, \bibinfo{pages}{63--88}.
\newblock


\end{thebibliography}

\end{document}